# Early Diagnosis of Lung Cancer Using Computer Aided Detection via Lung Segmentation Approach

Abhir Bhandary [#1], Ananth Prabhu G [*2], Mustafa Basthikodi [*3], Chaitra K M [*4]

[1]Assistant Professor, Department of Information Science Engineering, NMAM Institute of Technology, (Visvesvaraya Technological University, Belagavi), Nitte, Karnataka, India

[2,3] Professor, Department of Computer Science Engineering, Sahyadri College of Engineering and Management, (Visvesvaraya Technological University, Belagavi), Adyar, Karnataka, India

[4] Assistant Professor, Department of Computer Science Engineering, JSS Academy of Technical Education, Noida, India

[1] abhirbhandary@nitte.edu.in, [2] educatorananth@gmail.com, [3] mbasthik@gmail.com

*Abstract: Lung cancer begins in the lungs and leading to the reason of cancer demise amid population in the creation. According to the American Cancer Society, which estimates about 27% of the deaths because of cancer. In the early phase of its evolution, lung cancer does not cause any symptoms usually. Many of the patients have been diagnosed in a developed phase where symptoms become more prominent, that results in poor curative treatment and high mortality rate. Computer Aided Detection systems are used to achieve greater accuracies for the lung cancer diagnosis. In this research exertion, we proposed a novel methodology for lung Segmentation on the basis of Fuzzy C-Means Clustering, Adaptive Thresholding, and Segmentation of Active Contour Model. The experimental results are analysed and presented.*

**Keywords:** *Computer Aided Detection, Diagnosis, Fuzzy clustering, Lung cancer, Segmentation, Thresholding*

## I. INTRODUCTION

Cancer is the considered to be a primary reason for world's population demise, accounting for almost 8.2 million fatalities per annum. Out of which one million expiries per year is caused due to Lung cancer [1]. Early detection through LDCT scans showed a 16-20 percent decrease in lung cancer mortality compared to normal adult chest X-ray mortality. Unlike traditional X-rays, Low Dose Computer Tomography scanning gives very precise three-dimensional images of various types of tissue, preventing multiple levels of distinct tissues from being overlapped in a single image.

Lung cancer is a wicked tumour distinguished by hysterical growth cells in lung tissues [2]. Its progress can blowout into adjacent tissue or further fragments of bodily tissues outside the lung via a process called metastasis. Though most lung nodules are not cancerous, non-lung tissue detection involves the analysis of cancerous lung in order to be able to accomplish a surgery to make sure the cancerous property is benign. Thus, discovery of nodules is closely associated with the diagnosis of cancer. However, the compromise of assessing lung nodules by radiologists is set to fifty two percent when perceiving nodules of any extent.

A considerable amount of study has been carried out over the last twenty years in computer aided lung cancer detection systems in low dosage CT scans termed as LDCT. Many cancer detection systems are projected in the literature. However, there are issues with high false positive rates and low sensitivity which eliminate aforementioned systems from being utilized in day-to-day medical practice.

Lung cancer is another greatest prevalent form of cancer amid women and men, and this form of cancer is accountable for the death of around one third of people. Lung cancer has instigated several demises among women than any other carcinogenic disease in the established nations in the recent years.

Performing the diagnosis of lung cancer is a imperative factor in the transience and its delay can aggravate the mortality conditions. Common indicators of lung cancer include fatigue, chest pain, coughing, haemoptysis, sore throat, weight loss and chest infection. The need for efficiency as well as prompt application of therapeutic and diagnostic procedure is demonstrated in this. Multiple variables may lead to shortfalls in out-patient care for cancer patients in healthcare facilities: Accessibility to specialist medical care, patient referral problems & lack of particular diagnosis. Adequate patient handling within a public health system is necessary for speeding up counselling, respond to diagnostic test requests, & minimise the lag in tumour diagnosis and treatment.

Screening tests are suggested for individuals with high risk of contracting the disease like long-term exposure to carcinogen, coal miner's and long history of smoking. Sputum cytology, Chest X-ray and Computed Tomography have been studied for a long time as alternatives for lung cancer screening. LDCT has recently become the new normal for lung cancer screening that reduces the risk by twenty percent when associated with chest X-rays. A significant clinical observation in Computed Tomography images is the Lung nodule. The chance of a nodule being malignant is around 40 percent. Differentiating among nodule and pulmonary vessels is a difficult task as they share identical shape & intensity characteristics. The difficulty level of nodule detection process increases when the lung nodules are attached near the lung wall or with the blood vessel. Computerized nodule detection schemes have shown tremendous improvement in diagnostic accuracy of lung cancer identification.

The following steps are involved in a typical Computer Assisted Diagnostic method: interest detection region, pre-processing, lung parenchyma segmentation, nodules classification and feature extraction. Image processing is dealt in the first four steps and final classification step encounters the problem of pattern recognition. Sputum cytology, Chest X-ray and Computed Tomography have been studied for a long time as the lung cancer screening alternatives. LDCT has become standard for screening lung cancer. Thousands of high-risk individuals had enrolled for lung cancer screening in the National Lung Screening Trial.

Lung nodule is a small oval or circular shaped lesion with non-uniform textures and good contrast in the thoracic CT image with respect to lung parenchyma. The diameter of the lung nodules is less than 30 mm. The lesion with a diameter of greater than 30 mm is called pulmonary mass which is more likely to be a cancer than a nodule. However, the early detection process of lung cancer process that includes identification of nodules whose diameter is between 3 and 10 millimetre is significantly very important. The likelihood that a nodule could be malignant is around 40%; for a suspicious lung nodule, however, a follow up is essential.

Any portion of the lung parenchyma can contain Lung nodules. Based on their spatial location, lung nodules are considered into three types:

a. Juxta-pleural nodules: they are attached to the surface of pleural.

b. Juxta-vascular nodules: Connected with pulmonary vessels.

c. Isolated nodule: it is a distinct, well-marginated nodule that has no relation with other dense anatomical structures.

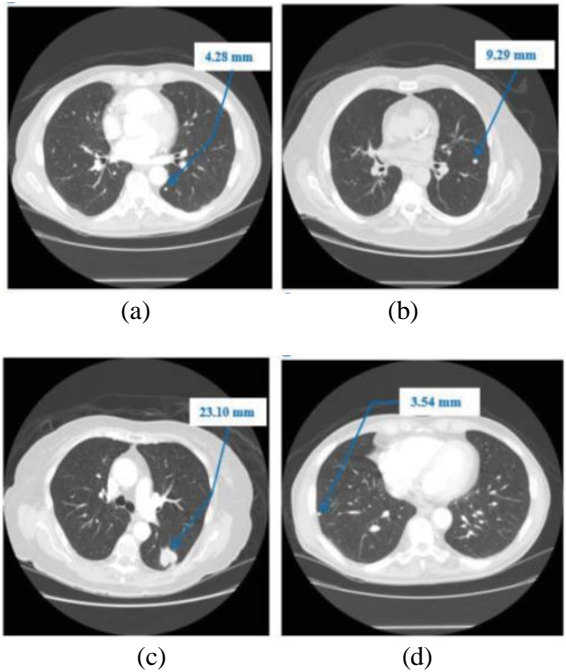

**Fig. 1 Isolated Nodules (a-b) and Juxta-Pleural Nodules (c-d)**

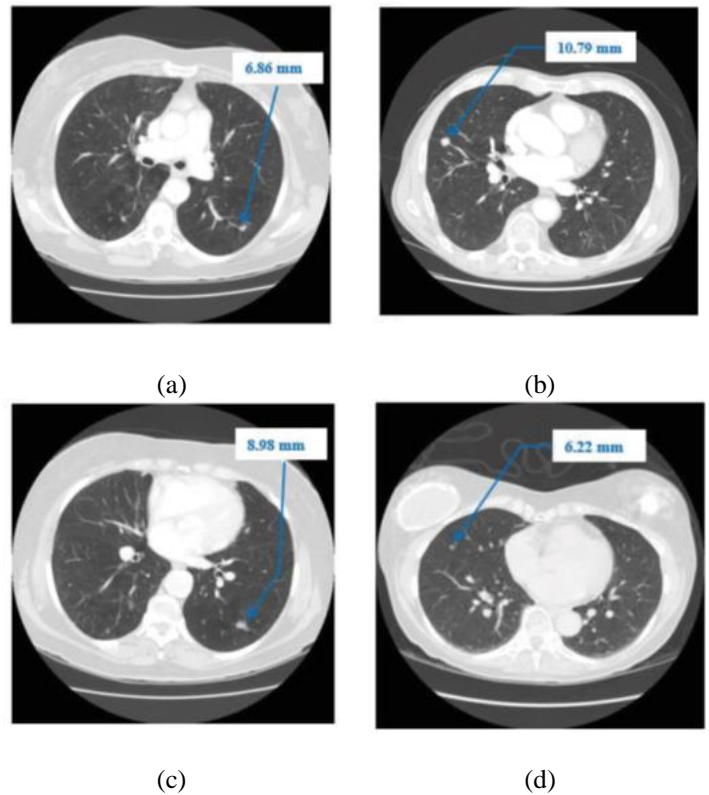

(a)          (b)

(c)          (d)

**Fig. 2 Juxta-Vascular Nodules (a-b); Part-Solid Nodule (c) Pure Ground Glass Nodule (d)**

## II. LITERATURE SURVEY

This section contains detailed study on existing methods that support the implementation of proposed technique. Cancer is a disease that is caused in any part of the human body by an uncontrolled eccentric cell division. Cancer is at the top in some places in the list of deadly diseases & widespread in every part of the world, but still raising. Premature exposure of lung cancer is cumbersome in several amount of cases. Efficient & effective means of technological lung cancer detection technique is presented in the research work in [3]. Authors utilized a collection of CT scanned images corresponding to lungs as input that were originally attained from records of lung image then implemented techniques such as segmentation, feature extraction for processing of image.

A research paper was presented at [4], this work consists of two varieties of cancer images that is CT & MRI. Using Gabor filter, scan images are being improved & Canny filter is introduced due to its precision for detection of edge & ultimately segmentation of super pixel is done. Full phases of medical image processing are addressed.

Researchers [5] [6] established a method using soft calculating and the various techniques pertaining to image processing systems for spotting the lung cancer from CT scanned photo images. Here, for pre-processing, anisotropic diffusion filter as well as Gaussian filter techniques happen to be applied. In the case of image segmentation, watershed segmentation algorithm is being applied which is more effective approach with strong success rate. SVM and KNN algorithms are used for classification & K Nearest Neighbour approach is showing a 5.5% greater classification outcome.

Writers [7] presented research paper that represents lung cancer edge detection techniques. Researchers show that Computed Tomography scan can detect lung nodules of size ranging between 2–3 mm that is very delicate and the mixture of watershed segmentation and Gabor filter pre-processing offers excellent outcomes. Outcome indicates Canny operator offers top grades for the purpose of edge detection when contrasted with other edge identification techniques.

Research in [8] resumed on lung tumour identification techniques, where Fast Fourier transform & Gabor filter techniques are compared but greater results are obtained by Gabor filter image enhancement method. Approaches like thresholding and Marker-Controlled Watershed Segmentation are analogized in the stage of image categorization. Around 4% greater results were produced by Watershed Segmentation method. Masking techniques & Binarization are matched for image feature extraction, but binarization method provides better results. Computer aided identification technique on the basis of hierarchical vector quantization scheme has been suggested by Authors [9] & it offers more precise segmentation compared to threshold method. SVM Classifier approach is utilized here, providing 92.7 percent sensitivity and 93.3 percent specificity and 82.7% overall sensitivity at 4FP/Scan. However, at 4.14 FPs/Scan specificity sensitivity of 89.3% is observed.

CT images are generally used for processing medical images, because of their low noise level and high resolution. The use of Computer-Aided identification system in lung cancer treatment is studied in this paper, consisting segmentation as well as pre-processing techniques, as well as data analysis methods [10]. In order to help with collection, analysis, evaluation of imagery medical information, the main objective was to discover new technology for computational diagnostic tools development. However, there are aspects which still need more attention, like increasing sensitivity, minimizing false positives, & maximize identification of every nodule type, even for various sizes and shapes. Researchers [11] described a general structure for identification of lung cancer in chest Low Dose Computer Tomography images. Our technique comprises of a detector nodule skilfully classified on LIDC-IDRI data base accompanied by a Kaggle DSB 2017 that was being trained by cancer predictor trained on data source.

The diagnosis of Computer-assisted lung cancer is split to 2 key issues, that is, lung cancer prediction & lung nodule detection. As a significant intermediate step for envisaging lung cancer, many algorithms have centred on identification of nodule [12] [13]. Some other methods aim to forecast nodules of cancer through nodule candidate patches, ignoring explicit nodules detection [2] [14]. An intermediate nodule detector is discovered by our system whose recognitions are afterward fed as input for predicting the cancer.

Detection of Nodule and/or classification beginning from subject Low Dose Computer Tomography has also been analysed via the techniques of Deep Learning. A popular approach is to use two-dimensional Convolutional Neural Network having orthogonal or persistent patches that are nodule–centred due to its forthright adaptation from typical CNN architectures organized for natural image classification and/or detection [15]. Furthermore, Low Dose Computer Tomography like several other biomedical image processing methods includes 3-dimensional data which real images do not contain and are not manipulated with 2-dimensional Convolutional Neural Network. A good technique is to make use of Convolutional Neural Network with 3-dimensional convolutions for reaping the benefits of 3-dimensional information. Various methods make use of 3D CNN for nodule detection, Categorization, or/and stratification [16] [17].

The research work undertaken [18] to analyse, review, classify & discuss the latest advancements in the identification of human body cancer through ML methods for liver, lung, breast, skin cancer leukaemia, brain. Research work illustrates the way in which ML with unsupervised, supervised and deep learning methods assists cancer diagnosis, cure processes. Various state of art methods is classified under similar cluster & outcomes are contrasted with benchmarked information from accuracy, specificity, sensitivity, false-positive metrics.

Authors [19] compared the lung cancer recognition algorithms & developed virtual machine images for maximum solutions. Though CNN obtained reasonable precision, there's a lot of place for enhancement in generalizability of models.

## III. MATERIALS AND METHODS

### A. Computer Aided Lung Cancer Diagnosis

The latest development in CT imaging technology vastly enhances scanning time & image resolution. With the progress of Multi-Detector Computed Tomography, more than 300 thin section images per thoracic CT examination may be obtained. This additional information assists the radiologists identify the abnormalities more precisely. Moreover, the pressure of this increasing workload inevitably falls on radiologist at the point of interpretation. The reported Computed Tomography interpretation error in routine radiology practice, is 3% to 4% and even greater (30%) in studies involving abnormalities.

The technologies developed to minimize observational oversights and enhance the diagnostic accuracy of radiologists are Computer Aided Detection and Computer Aided Diagnosis. In screening Computed Tomography, Computer Aided Detection may assist the radiologists avoid overlooking a lung nodule, whereas Computer Aided Diagnosis may help the radiologists determine whether nodule is malignant or benign. An improvement in nodule detection accuracy with Computer Aided Detection Assistance has been demonstrated in Retrospective clinical trials.

For lung cancer diagnosis, there is a huge variation in Computer Aided Diagnosis system. Figure 3 demonstrates a typical Computer Aided Diagnosis system; it comprises of the following stages: pre-processing, lung parenchyma segmentation, interest detection region, classification of nodule and feature extraction. The image processing id dealt in the first 4 stages & final classification step addresses the pattern recognition problem. Various techniques are being suggested in the literature to address every step but the issue is wide open & yet to be improved [20].

Very high sensitivity and specificity are the two main requirements of a Computer Aided Diagnosis system. Sensitivity is the Computer Aided Diagnosis system's ability to identify all nodules in the image. It may be enhanced by decreasing the false negatives (nodules which are not detected by system). The specificity is the capacity of the Computer Aided Diagnosis system to eliminate all the non-nodule structures in the image.

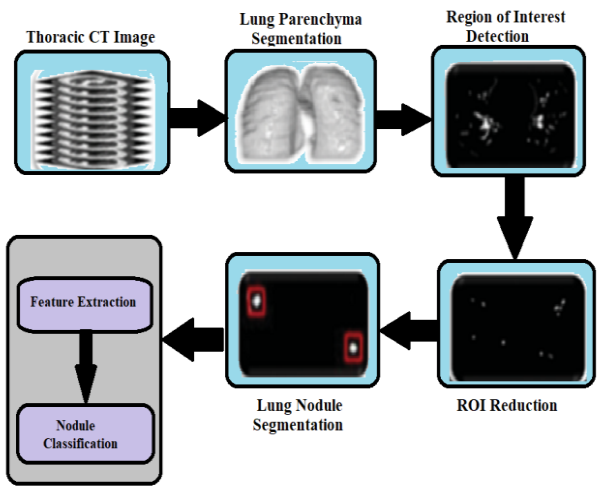

**Fig. 3 Lung Cancer Diagnosis System for Computer Aided Diagnosis**

By decreasing the false positives, it could be improved (system reports a non-nodule structure as a nodule). It is very essential to Keep the sensitivity very high (100%) since a single nodule missed matters for somebody's life & death. The high specificity value decreases the time spent by the physician on the non-nodule systems. Eventually, the system's computing time should be relatively low [21]. Various research works have showed that Computer Aided Diagnosis systems are very efficient in identifying minute pulmonary nodules in the CT images. Very little number of nodules found by screening Computer Tomography scans, however, are malignant that enhances the number of false positive detection. Because the blood vessels look very similar to lung nodules when looking at a single slice, all the slices need to be taken into consideration. Radiologist must start with a small region at the top of a lung, and move down the slices, searching for round structures that last at the same location for few slices. Similar steps must be followed till all parts of the lungs are considered. In order to address these drawbacks of current Computer Aided Diagnosis systems, false positives must be minimised by adding Region of Interest reduction measure that reduces the amount of ROI's to be processed. Besides that, nodule characterization must be integrated into the Computer Aided Diagnosis system to obtain high specificity rate and sensitivity [22].

The main focus of this research work is on multiple investigations on computer supported diagnostic framework for early detection of lung cancer from low dose computed tomographic images. The lung parenchyma segmentation from other anatomical structures in Low Dose Computed Tomography image decreases the difficulty of the computer aided nodule detection method. The size of each slice in the Low Dose Computed Tomography image is 512×512 pixels and size of both the lungs spans approximately one fourth of the slice size. The initial lung segmentation stage further enhances the computational speed & efficiency of subsequent processing steps. The accuracy of the segmentation of the lung parenchyma is also very significant since these lung nodules may exist at the lung walls.

### B. Lung Segmentation Approaches
With improvements as modelled in Figure 4, three Lung parenchyma segmentation techniques based on Adaptive Thresholding, Active Contour Model Segmentation and Fuzzy C-Means Clustering are suggested in this section.

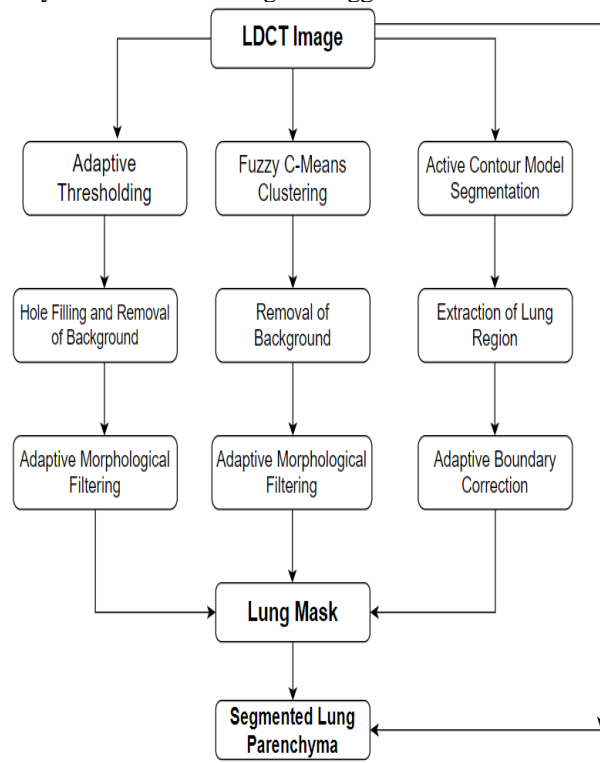

**Fig. 4 Proposed Block Diagram of Lung Segmentation on the basis of Adaptive Thresholding, Fuzzy C-Means Clustering & Active Contour Model Segmentation.**

### a) Adaptive Thresholding
Application of thresholding for the segmentation of lung parenchyma & potential improvements in the segmentation accuracy through Adaptive Morphological Filtering technique is suggested. Thresholding was used as the primary segmentation technique by most of the lung field segmentation algorithms and error introduced by thresholding procedure was later improved. Figure 5 shows the limitations of thresholding-based segmentation.

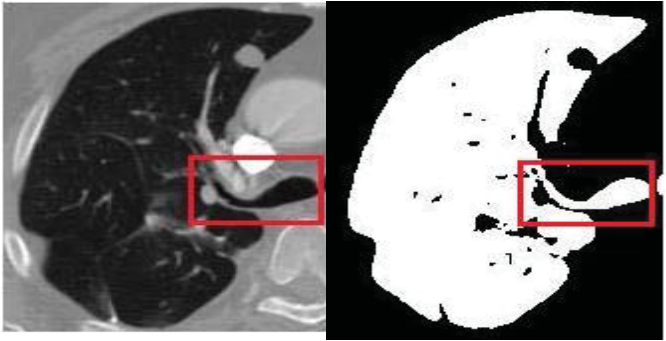

**(a) Inclusion of Airways**

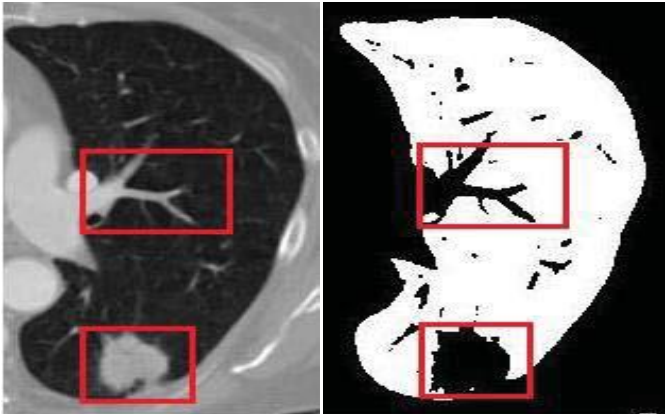

**(b) Exclusion of pleural nodules and pulmonary vessels**

**Fig. 5 Limitations of Segmentation Strategies of Gray Level Thresholding**

Adaptive thresholding of the Otsu is used along with morphological opening as well as closing to lung parenchyma segment suggested by [23]. Morphological filtering with fixed size structuring element, however, will cause major edge distortions because of lung boundary smoothing, which decreases the accuracy of segmentation. The bi-level thresholding of Otsu is implemented to Low Dose Computed Tomography image in such a way that lung-parenchyma is represented by white region in the output binary image. The thresholding mechanism is just reversed in this case and represented by

$$st = \begin{cases} Lt & when \quad F \leq T \\ 0 & when \quad F > T \end{cases} \qquad [1]$$

The nodules & blood vessels appear as dark holes in the lung parenchyma. For creating the lung mask, these dark holes should be filled with the white pixels. For removing boundary objects and filling holes within the lung area, an adaptive morphological filtering method is suggested. In this approach, an adaptive structuring element is added, which has pre-defined shape, but the size varies according to the characteristics of local boundary. After creating the lung mask, the multiplication of lung mask with original LDCT image results in the segmentation of the lung area. Figure 6 shows the outcome of lung segmentation & the performance at different intermediate stages.
The nodules & blood vessels appear as dark holes in the lung parenchyma. For creating the lung mask, these dark holes should be filled with the white pixels. For removing boundary objects and filling holes within the lung area, an adaptive morphological filtering method is suggested. In this approach, an adaptive structuring element is added, which has pre-defined shape, but the size varies according to the characteristics of local boundary. After creating the lung mask, the multiplication of lung mask with original LDCT image results in the segmentation of the lung area. Figure 6 shows the outcome of lung segmentation & the performance at different intermediate stages.

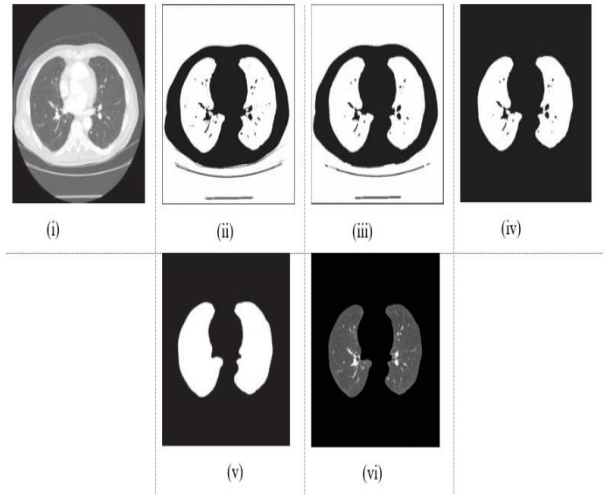

**Fig. 6 Segmentation of Lungs Using Adaptive Border Correction and Thresholding. (i) Low Dose Computed Tomography Image; (ii) Reverse Thresholding (iii) Median filtering (iv) Background removal (v) Lung Mask after AMF (vi) Segmented Lung Parenchyma**

*b) Fuzzy C-Means Clustering*
Here, for accurate delineation of lung parenchyma, we suggest an unsupervised segmentation technique on the basis of fuzzy c-means clustering followed by mask processing and lung border refinement. Conventional clustering algorithm allocate only one cluster to each data point. Fuzzy clustering, on the other hand, renders data points to belong to more than one cluster depending on degree of the membership feature connected with each cluster. In this case, the cluster centre and the membership function are initialized randomly & iteratively researches adapted fuzzy c-means approach to find optimal threshold for segmenting the lung parenchyma. When compared with k-means algorithm, the performance of Fuzzy c-Means Clustering in case overlapping data was much better.
Figure 4 depicts the suggested fuzzy c-means clusteri ng for lung-parenchyma segmentation. Fuzzy c-means clustering performs initial lung segmentation through the Low Dose Computed Tomography image, the outcome of the segmentation is better when compared to thresholding technique. However, the lung boundary abnormalities and nodules and the presence of pulmonary vessels need further processing to develop lung mask.
Figure 7 depicts the results of lung-parenchyma segmentation through adaptive morphological filtering & fuzzy c-means grouping. The result of the segmentation of Fuzzy C-means Clustering is shown in Figure 7(ii). Figure 7(v) describes the development of lung mask through Adaptive Morphological Filtering. Figure 7(vi) illustrates that the lung mask is multiplied with the original Low Dose Computed Tomography image to get the lung parenchyma.

### c) Active Contour Modelling

This technique makes use of the image global data to find the stopping criteria. The introduction of the global energy term reduces the mistake introduced by noise in the image. The block diagram in Figure 4 depicts the lung parenchyma segmentation scheme of the proposed Active Contour Model. This approach includes Active Contour Model segmentation followed by refinement process of adaptive morphological boundaries. As the lung field is directly segmented by the application of active contour technique, this approach does not require border removal and region filling steps. This technique does not, however, involve juxta-pleural nodules, as shown in Figure 8(iii). In order to include juxta-pleural nodule in the lung area, lung boundaries are corrected using the suggested method. Figure 8 provides the results of various stages of the proposed technique.

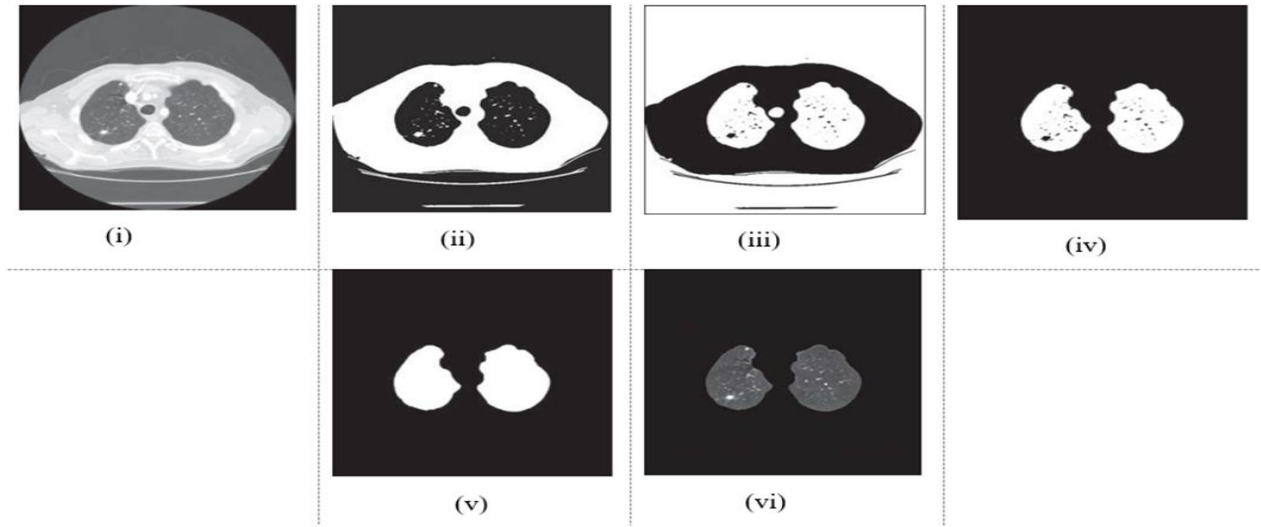

**Fig. 7 Segmentation of Fuzzy c-means: (i) Low Dose Computed Tomography Image (ii) Fuzzy C-means Clustering (iii) Image negatives (iv) Removal of surrounding tissues (v) Lung mask after Adaptive Morphological Filtering (vi) Segmented Lung-Parenchyma**

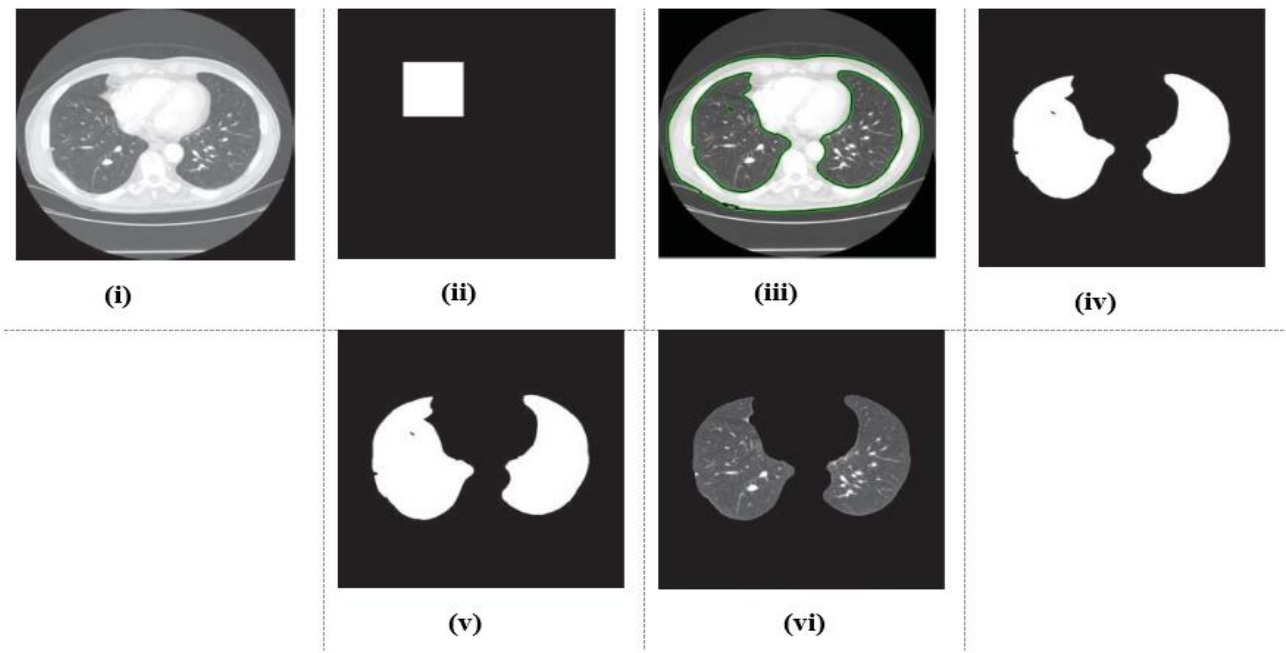

**Fig. 8 Segmentation of Region-Based Active Contour: (i) Low Dose Computed Tomography Image (ii) Initial region selection (iii) Active Contour Segmentation (iv) Separation of Lung region (v) Lung mask after Adaptive Morphological Filtering (vi) Segmented Lung-Parenchyma**

## C. Performance Evaluation Metrics

We have imposed four metrics for testing the efficiency of suggested method: Hausdorff Distance, Specificity, Dice Similarity Coefficient and Sensitivity. The overall amount of each combination of pixels for both automatic and manually labelled image is defined as below:

True Positive indicates the total number of pixels that are properly labelled by the algorithm. The amount of lung parenchymal pixels that the algorithm failed to label is indicated as False Negative. False Positive describes the amount of non-lung parenchymal pixel classified as lung parenchyma by the segmentation algorithm. The number of pixels which are not labelled as lung parenchyma by both manual and automatic segmentation is represented by (TN) True Negative.

Dice Similarity Coefficient is a statistical metric that evaluates the degree of similarity among the manual & automatic segmented images. The DSC is defined as below:

$$DSC = \frac{2TP}{2TP+FP+FN} \qquad [2]$$

Dice Similarity Coefficient may take values between 0 & 1. Dice Similarity Coefficient will be equal to 1 if both automatic and manual segmentation results are identical and vice versa. Specificity and sensitivity are statistical techniques of the algorithm's ability to correctly label the lung parenchymal pixels as well as correctly exclude not a lung parenchymal pixel respectively [24]

$$Specificity = \frac{TN}{TN+FP} \qquad [3]$$

$$Sensitivity = \frac{TP}{TP+FP} \qquad [4]$$

A measurement of error between the two segmented regions is provided by the Hausdorff distance. Let Sx & Sy be two surfaces, then the Hausdorff symmetrical distance is defined as:

$$D_H (S_x, S_y) = Max [D (S_x, S_y), D (S_y, S_x)] \qquad [5]$$

where,

$D (S_x, S_y) = Max [D (m, S_y)], m \varepsilon S_x$
$D (S_y, S_x) = Max [D (m, S_x)], m \varepsilon S_y$

## IV. RESULTS AND DISCUSSION

The proposed approach has been tested by making use of database created by Lung Image Database Consortium. The data base comprises of 119 normal and 70 abnormal lung Computerized Tomography images from 28 cases. The eligibility criteria for abnormal image includes the following condition: Must have 40% of pathological lung image that includes pleural nodules & 20 percent of the data comprises of pulmonary vessels and airways connected with lung walls. With the help of experts, the ground truth images are carefully labelled. The output of segmentation schemes is analysed in terms of its specificity values, sensitivity, Dice Similarity Coefficient and Hausdorff Distance.

The efficiency of numerous lung-parenchyma segmentation schemes like combination of thresholding, Fuzzy C-means Clustering, rolling ball algorithm and region based Active Contour Segmentation was calculated. The boundary correction phase in the current lung segmentation techniques is replaced with the proposed border correction technique to determine the robustness of the adaptive border correction technique. Later, its performance is being compared with the actual system performance. The findings show that the suggested border correction technique improve the accuracy of segmentation. Table-1 represents the Dice Similarity Coefficient and the Hausdorff Distance of various segmentation techniques.

**Table-1: Comparison of performance of Adaptive Border Correction Technique in multiple Lung Segmentation Techniques**.

| Segmentation Approaches | Active Contour Model | | Fuzzy C-Means | | Adaptive Thresholding | |
|---|---|---|---|---|---|---|
| Correction of Boundary | Proposed | GMM | Proposed | Morphological Filtering | Proposed | Rolling Ball Algorithm |
| Hausdorff Distance (HD) | 1.27 | 1.53 | 1.52 | 1.78 | 1.95 | 2.15 |
| Dice Similarity Coefficient (%) | 97.14 | 93.83 | 94.01 | 91.41 | 93.29 | 88.07 |

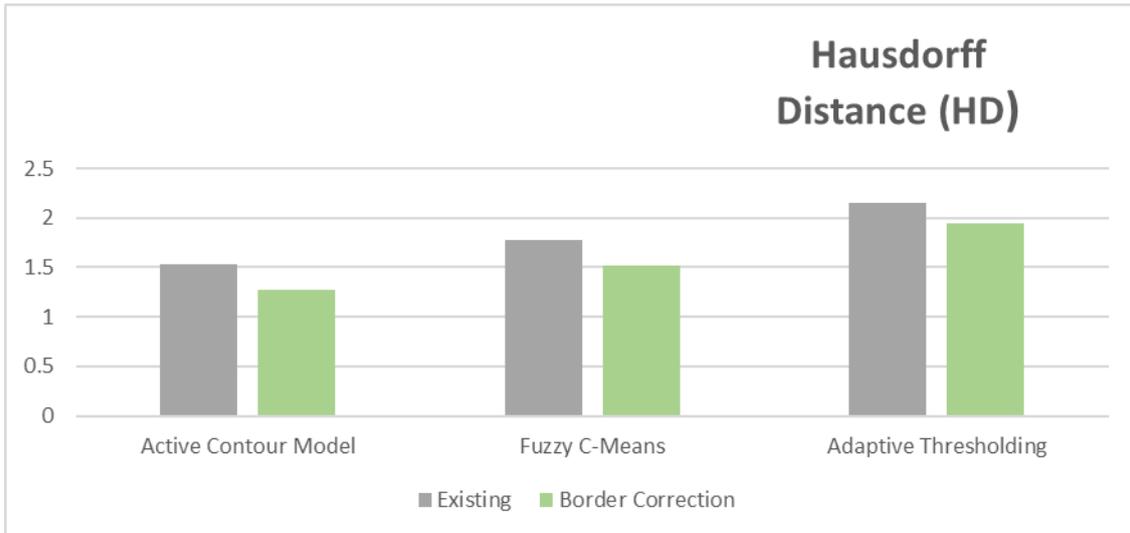

**Fig. 9 Hausdorff Distance of Various approaches for Lung Segmentation**

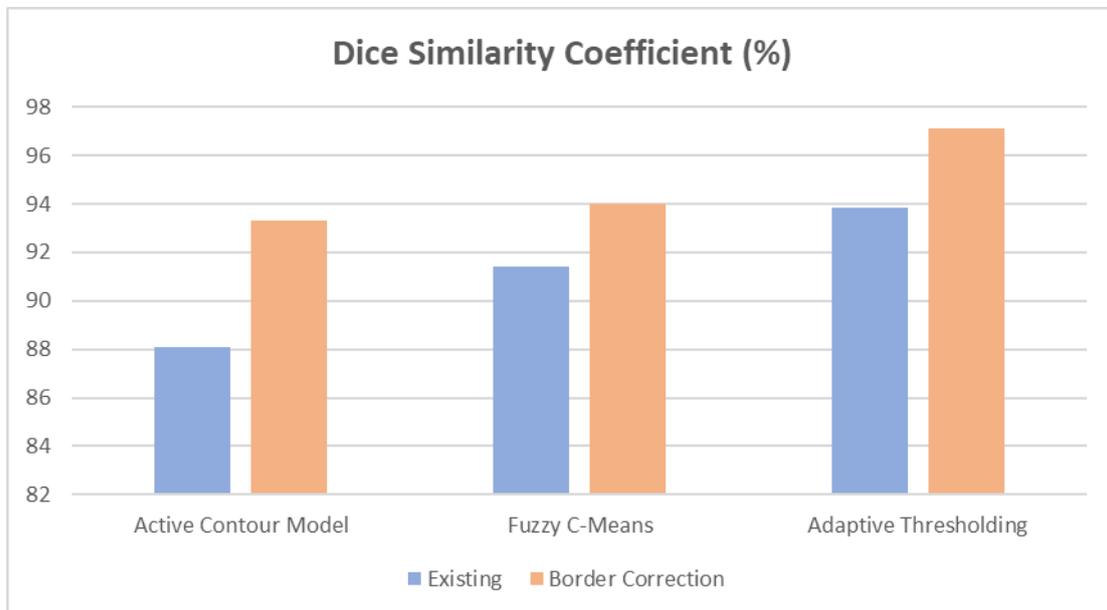

**Fig. 10 Dice similarity Coefficients of different methods of Lung Segmentation**

Figure 9 and Figure 10 graphically represents the improvements in the segmentation outcome with the introduction of the proposed border correction strategy. The outcomes show that there is an improvement in the existing systems performance. There seems to be an average of 4% to 5% increase of DSC & approximately 20% to 25% of decline in Hausdorff Distance.

## V. CONCLUSIONS

The Computer Aided Diagnosis System for Lung Cancer Diagnosis is discussed. An improved Lung parenchyma segmentation technique based on Adaptive Thresholding, Active Contour Model Segmentation and Fuzzy C-Means Clustering are suggested in the paper. The proposed scheme does not involve interventions of human, which is unsupervised GMM dependent segmentation followed by an adaptive border correction approach, in order to reduce computation time. The experimental results demonstrate the improved accuracy of segmentation over other approaches. This approach would be a promising approach for lung cancer-based computer aided diagnosis system.

## VI. ACKNOWLEDGEMENT

Our appreciation goes to everyone who provided guidance and support to this study.